\documentclass[jfm,aps,preprint,showpacs,preprintnumbers,amsmath,amssymb,secnumroman,eqsecnum]{revtex4}

\usepackage{graphicx}
\usepackage{dcolumn}
\usepackage{bm}

\newcommand{\beq}{\begin{equation}}
\newcommand{\eeq}{\end{equation}}
\newcommand{\beqa}{\begin{eqnarray}}
\newcommand{\eeqa}{\end{eqnarray}}
\newcommand{\vc}[1]{\mbox{\boldmath $#1$}}

\newcommand{\vol}[1]{{\bf #1}}

\newcommand{\du}[1]{{\bf\sf #1}}

\begin{document}


\title{Coaxial collisions of a vortex ring and a sphere in an inviscid incompressible fluid}

\author{B. U. Felderhof}

 \email{ufelder@physik.rwth-aachen.de}
\affiliation{Institut f\"ur Theorie der Statistischen Physik\\ RWTH Aachen University\\
Templergraben 55\\52056 Aachen\\ Germany\\
}%

\date{\today}

\begin{abstract}
The dynamics of a circular thin vortex ring and a sphere moving along the symmetry axis of the ring in an inviscid incompressible fluid is studied on the basis of Euler's equations of motion. The equations of motion for position and radius of the vortex ring and those for position and velocity of the sphere are coupled by hydrodynamic interactions. The equations are cast in Hamiltonian form, from which it is seen that total energy and momentum are conserved. The four Hamiltonian equations of motion are solved numerically for a variety of initial conditions.
\end{abstract}

\maketitle
\section{\label{I}Introduction}
The flow pattern of a circular line vortex ring, moving by itself in an inviscid incompressible fluid, was first studied by Helmholtz \cite{1}. The theory is reviewed in the monographs of Lamb \cite{2} and Batchelor \cite{3} and in the articles by Shariff and Leonard \cite{4} and Meleshko \cite{5}. Multiple coaxial vortex rings exhibit fascinating interactions \cite{5}. The interaction of a single vortex ring with a fixed sphere centered on its axis was first studied by Dyson \cite{6}. In the following we study a single circular vortex ring interacting with a sphere which can move freely along the axis of the ring. When the ring and sphere are widely separated they move independently. Their interaction may be viewed as a collision.

The flow pattern of the fluid at any instant of time can be decomposed into irrotational and rotational contributions. For a fixed sphere the flow pattern is purely rotational and consists of a Helmholtz vortex ring and its image in the sphere, as first found by Lewis \cite{7}. A moving sphere by itself carries an irrotational dipolar flow pattern. If both ring and sphere are present, then the image ring and the dipolar pattern of the sphere act on the vortex ring. Thus the coaxial sphere and vortex ring constitute a complicated dynamical system. In order to describe its motion we formulate Hamiltonian equations of motion. Different types of motion can be analyzed by numerical solution of the four Hamiltonian equations. There are two constants of the motion, the total energy and the total momentum.

We find that there are several distinct types of motion. Depending on the initial conditions the vortex ring can pass over the sphere, widening as it does so. Alternatively, the sphere can pass through the ring and leave it behind. In the latter case the ring also widens during the passing. There are also situations, where the ring and sphere do not pass, but both slow down as a result of the encounter.

It would be of interest to study these motions experimentally. We consider an idealized system where the intrinsic instability and the interaction-induced instability of the vortex ring \cite{8} are ignored. In a superfluid vortex rings are particularly stable \cite{9}. The motions in such a fluid would be worthy of investigation.

\section{\label{2}Flow patterns and self-energy}

We consider a sphere of radius $b$ immersed in an inviscid incompressible fluid of mass density $\rho$. The fluid flow velocity $\vc{v}(\vc{r},t)$ and the pressure $p(\vc{r},t)$ satisfy Euler's equations of motion
\begin{equation}
\label{2.1}\rho\big[\frac{\partial\vc{v}}{\partial t}+(\vc{v}\cdot\nabla)\vc{v}\big]=-\nabla p,\qquad\nabla\cdot\vc{v}=0.
\end{equation}
The sphere is impenetrable to the flow.

The vorticity of the flow is defined as
\begin{equation}
\label{2.2}\vc{\omega}=\nabla\times\vc{v}.
\end{equation}
The flow velocity at any time $t$ can be expressed as the sum of an irrotational and a rotational flow
\begin{equation}
\label{2.3}\vc{v}=-\nabla\phi+\nabla\times\vc{A},
\end{equation}
with scalar potential $\phi$ and vector potential $\vc{A}$. The potentials satisfy the equations
\begin{equation}
\label{2.4}\nabla^2\phi=0,\qquad\nabla^2\vc{A}=-\vc{\omega},\qquad\nabla\cdot\vc{A}=0.
\end{equation}
The last equation represents a convenient choice of gauge. In the following we do not need to evaluate the pressure.

If the sphere is centered at the origin and moves with translational velocity $\vc{U}$ in infinite fluid then it generates the scalar potential \cite{10}
\begin{equation}
\label{2.5}\phi(\vc{r})=\frac{1}{2}\;b^3\frac{\vc{r}}{r^3}\cdot\vc{U},\qquad r>b,
\end{equation}
corresponding to the dipole moment
 \begin{equation}
\label{2.6}\vc{q}=\frac{1}{2}\;b^3\vc{U}.
\end{equation}
The corresponding Poisson flow pattern $\vc{v}=-\nabla\phi$ is given by
\begin{equation}
\label{2.7}\vc{v}(\vc{r})=\du{F}(\vc{r})\cdot\vc{q},\qquad \du{F}(\vc{r})=\frac{-\du{I}+3\hat{\vc{r}}\hat{\vc{r}}}{r^3},
\end{equation}
where $\du{F}(\vc{r})$ is the dipolar tensor with unit tensor $\du{I}$.
The dipole moment is related to the impulse $\vc{S}$ which is required to accelerate the sphere from rest to velocity $\vc{U}$ by
\begin{equation}
\label{2.8}\vc{q}=\beta\vc{S},\qquad\beta=\frac{b^3}{2m^*},
\end{equation}
where $m^*$ is the effective mass
\begin{equation}
\label{2.9}m^*=m+\frac{1}{2}m_{f},\qquad m_{f}=\frac{4\pi}{3}\rho b^3.
\end{equation}
Here $m_{f}$ is the mass of displaced fluid and $\frac{1}{2}m_{f}$ is the added mass. The kinetic energy of the flow is evaluated to be \cite{10}
\begin{equation}
\label{2.10}K_s=\frac{1}{2}\rho\int_{r>b}\vc{v}^2\;d\vc{r}=\frac{1}{2}\rho\int_S\phi\vc{U}\cdot\vc{n}\;dS=\frac{1}{4}m_f\vc{U}^2,
\end{equation}
where $\vc{n}$ is the outward normal to the spherical surface $S$.

Next we consider a single vortex ring centered at the origin in the absence of the sphere and moving along the $z$ axis. On account of axial symmetry it is convenient to use cylindrical coordinates $(s,\varphi,z)$. Then the vector potential can be expressed as \cite{3}
\begin{equation}
\label{2.11}\vc{A}(\vc{r})=\psi(s,z)\vc{e}_\varphi,
\end{equation}
with stream function $\psi(s,z)$ and azimuthal unit vector $\vc{e}_\varphi$. For a circular line vortex of radius $R$ and strength $\Gamma$ the stream function is given by \cite{2},\cite{3}
\begin{equation}
\label{2.12}\psi(s,z,R)=\Gamma\frac{\sqrt{Rs}}{2\pi}\bigg[\bigg(\frac{2}{k}-k\bigg)K(k^2)-\frac{2}{k}E(k^2)\bigg],
\end{equation}
with variable {k} given by
\begin{equation}
\label{2.13}k^2=\frac{4Rs}{z^2+(s+R)^2},
\end{equation}
and complete elliptic integrals $K(k^2)$ and $E(k^2)$ in the notation of Abramowitz and Stegun \cite{11}. The kinetic energy of the corresponding flow diverges due the assumption that the vorticity is concentrated in a line. If the line element is expanded into a cylindrical filament of radius $a$ with uniform vorticity then for $a<<R$ the kinetic energy of the ring vortex is calculated to be \cite{2},\cite{12},\cite{13}
\begin{equation}
\label{2.14}K_r=\frac{1}{2}\rho\int\vc{v}^2\;d\vc{r}=\rho\int\vc{\omega}\cdot\vc{A}\;d\vc{r}=\frac{1}{2}\rho\Gamma^2R\bigg[\log\frac{8R}{a}-\frac{7}{4}\bigg].
\end{equation}
This shows that the energy diverges logarithmically as $a\rightarrow 0$. The ring moves in the axial direction with constant radius $R$ and velocity given by \cite{2}
\begin{equation}
\label{2.15}R\frac{dZ}{dt}=\frac{\Gamma}{4\pi}\bigg[\log{\frac{8R}{a}-\frac{1}{4}}\bigg],
\end{equation}
where $Z$ is the $z$ coordinate of the ring. By convention $\Gamma$ is taken to be positive, so that the ring moves in the positive $z$ direction.

We must take into account that according to Helmholtz's vortex equations \cite{2},\cite{10} the core radius $a$ shrinks as $a=wb\sqrt{b/R}$ with dimensionless constant $w$, as the radius of the ring increases, implying that the volume of the ring is constant. Here we adopted the radius $b$ of the sphere as the basic length scale. The equations of motion can be put in the Hamiltonian form
\begin{equation}
\label{2.16}\frac{dZ}{dt}=\frac{\partial\mathcal{H}_r}{\partial P_r},\qquad \frac{dP_r}{dt}=-\frac{\partial\mathcal{H}_r}{\partial Z},\qquad P_r=\pi\rho\Gamma R^2,
\end{equation}
with ring Hamiltonian given by
\begin{equation}
\label{2.17}\mathcal{H}_r(P_r)=\frac{1}{2}\rho\Gamma^2\sqrt{\frac{P_r}{\pi\rho\Gamma}}\bigg[\frac{3}{4}\log\frac{P_r}{\pi\rho\Gamma b^2}-\frac{7}{4}+\log\frac{8}{w}\bigg].
\end{equation}
 The ring Hamiltonian takes the value $K_r$ given by Eq. (2.14) with $a=wb\sqrt{b/R}$ for ring momentum $P_r=\pi\rho\Gamma R^2$. For an isolated ring the ring momentum is constant, since $\mathcal{H}_r$ does not depend on the position $Z$.

\section{\label{III}Sphere and ring}

If both the sphere and the ring are present, then they interact. We assume that both are centered on the $z$ axis with the ring parallel to the $xy$ plane. We denote the $z$ coordinate of the sphere center as $z_0$. In Fig. 1 we present a sketch of the geometry.

The scalar potential is given by Eq. (2.5) with $\vc{r}$ replaced by $\vc{r}-z_0\vc{e}_z$. The vector potential is no longer given by Eqs. (2.11) and (2.12) with $\vc{r}$ replaced by $\vc{r}-Z\vc{e}_z$, since we must take account of the kinematic boundary condition that the flow must be tangential to the sphere in its rest system. The total stream function is given by the above expressions plus the corresponding one of an image ring located inside the sphere \cite{7}. The total stream function is given by
 \begin{equation}
\label{3.1}\Psi(s,z,Z,R,z_0,b)=\psi(s,z-Z,R)-\sqrt{\frac{R}{R^*}}\;\psi(s,z-Z^*,R^*),
\end{equation}
with image coordinates $R^*,Z^*$ given by
\begin{equation}
\label{3.2}R^*=\frac{b^2R}{R^2+(Z-z_0)^2},\qquad Z^*=z_0+\frac{b^2(Z-z_0)}{R^2+(Z-z_0)^2}.
\end{equation}
The kinetic energy of the corresponding flow pattern is given by
\begin{equation}
\label{3.3}K_{rr^*}=K_r-\pi\rho\Gamma\sqrt{\frac{R}{R^*}}\;\psi(R,Z-Z^*,R^*),
\end{equation}
where the first term is the self-energy made finite, given by Eq. (2.14). The second term represents the interaction energy of the vortex ring with its image in the sphere. For large distance of sphere and ring the second term decays as $|Z-Z^*|^{-3}$, so that $K_{rr^*}$ tends to $K_r$. The expression may be compared with  Dyson's energy equation \cite{6}. The left-hand side of his equation (117) includes the self-energy of the image ring, which equals that of the ring by itself, and also includes the interaction energy twice. The terms in his energy equation are therefore twice those given in Eq. (3.3). In present notation Dyson's energy equation reads $2K_{rr^*}=2K_{r}(R_\infty)$, where $R_\infty$ is the radius of the ring for infinite separation from the sphere, as determined from the equation.

In order to find the total energy of sphere and ring we must take account of the fact that the ring moves also due to the dipolar flow pattern given by Eq. (2.7). According to Helmholtz the vorticity is carried along by the flow. As a consequence $R$ and $Z$ change according to
\begin{equation}
\label{3.4}\frac{dR}{dt}=v_s\big|_{R,Z},\qquad \frac{dZ}{dt}=v_z\big|_{R,Z},
\end{equation}
with $\vc{v}$ calculated from Eq. (2.7). This yields for the rate of change of the ring coordinates due to the potential flow pattern of a sphere of radius $b$ centered at $z_0$ with velocity $U$
\begin{eqnarray}
\label{3.5}\frac{dR}{dt}&=&\frac{3R(Z-z_0)}{[R^2+(Z-z_0)^2]^{5/2}}\;\frac{b^3}{2}U,\nonumber\\ \frac{dZ}{dt}&=&\frac{-R^2+2(Z-z_0)^2}{[R^2+(Z-z_0)^2]^{5/2}}\;\frac{b^3}{2}U.
\end{eqnarray}
Introducing the ring momentum $P_r=\pi\rho\Gamma R^2$ we can rewrite these equations in Hamiltonian form
\begin{equation}
\label{3.6}\frac{dZ}{dt}=\frac{\partial V_{rs}}{\partial P_r},\qquad\frac{dP_r}{dt}=-\frac{\partial V_{rs}}{\partial Z},
\end{equation}
with ring-sphere interaction
\begin{equation}
\label{3.7}V_{rs}=A_{rs}U,\qquad A_{rs}=\frac{b^3P_r}{[P_r/(\pi\rho\Gamma)+(Z-z_0)^2]^{3/2}}.
\end{equation}

Combining the above expressions we formulate the complete equations of motion in Hamiltonian form
\begin{eqnarray}
\label{3.8}\frac{dZ}{dt}&=&\frac{\partial \mathcal{H}_{rs}}{\partial P_r},\qquad\frac{dP_r}{dt}=-\frac{\partial \mathcal{H}_{rs}}{\partial Z},\nonumber\\
\frac{dz_0}{dt}&=&\frac{\partial \mathcal{H}_{rs}}{\partial p_0},\qquad\frac{dp_0}{dt}=-\frac{\partial \mathcal{H}_{rs}}{\partial z_0},
\end{eqnarray}
with Hamiltonian
\begin{equation}
\label{3.9}\mathcal{H}_{rs}(Z,P_r,z_0,p_0)=\mathcal{H}_r(P_r)-\pi\rho\Gamma\sqrt{\frac{R}{R^*}}\;\psi(R,Z-Z^*,R^*)+\frac{1}{2m^*}\big(p_0+A_{rs}(Z,P_r,z_0)\big)^2,
\end{equation}
where in the second term Eq. (3.2) must be used and subsequently $R$ must be replaced by $\sqrt{P_r/(\pi\rho\Gamma)}$. Since the Hamiltonian does not depend explicitly on time, energy is conserved and given by the value of $\mathcal{H}_{rs}$ at the phase point $(Z,P_r,z_0,p_0)$ considered. Since the Hamiltonian depends on coordinates $Z,z_0$ only via the difference $Z-z_0$ it is translation-invariant and total momentum
\begin{equation}
\label{3.10}P=P_r+p_0
\end{equation}
is also conserved. Hence we can reduce the problem to motion in a two-dimensional phase space with phase points $(r,P_r)$, where $r=Z-z_0$ is the relative position of ring and sphere. The equations of motion of the sphere-ring complex reduce to
\begin{equation}
\label{3.11}\frac{dr}{dt}=\frac{\partial\mathcal{H}_{red}}{\partial P_r},\qquad\frac{dP_r}{dt}=-\frac{\partial\mathcal{H}_{red}}{\partial r},
\end{equation}
where the reduced Hamiltonian $\mathcal{H}_{red}(r,P_r)$ is obtained from $\mathcal{H}_{rs}$ by substitution of $z_0=Z-r$ and $p_0=P-P_r$. The reduced Hamiltonian depends parametrically on total momentum $P$. Energy is again conserved and given by the value of $\mathcal{H}_{red}$ at the phase point $(r,P_r)$ considered.

\section{\label{IV}Motions of ring and sphere}

The equations of motion derived above are complicated, but can be easily solved numerically, either in the form of Eq. (3.8) or (3.11). We can think of Eq. (3.8) as an initial value problem to be solved for given initial values $Z(0),P_r(0),z_0(0),p_0(0)$ at time $t=0$. As noted above, by convention $\Gamma>0$, so that for large relative distance the ring moves in the positive $z$ direction. We can choose the initial position of the sphere at the origin, so that $z_0(0)=0$. Then there are several classes of motion, depending on the initial position $Z(0)$ of the ring, and the values of the initial momenta $P_r(0)$ and $p_0(0)$.

It is instructive to solve Eq. (3.8) for various initial conditions. The results can be compared with the solution of Eq. (3.11), which contains less information. In the following we consider some typical cases. In all cases studied below we take the sphere to be neutrally buoyant and centered at the origin at time $t=0$. We use vortex width parameter $w=0.01$.

In Figs. 2 and 3 we consider a collision, where initially the sphere is at rest and the ring is approaching from the left. The initial position of the ring is $Z(0)=-5b$ and the initial momenta are $P_r(0)=10\rho\Gamma b^2$ and $p_0(0)=0$. The ring passes over the sphere. In Fig. 2 we show the reduced positions $Z(\tau)/b$ and $z_0(\tau)/b$ as functions of reduced time $\tau=\Gamma t/b^2$. In Fig. 3 we show the momentum $P_r(\tau)$ in units $\rho\Gamma b^2$ as a function of $\tau$. We recall that the momentum of the ring is related to its radius $R$ by $P_r=\pi\rho\Gamma R^2$, so that the plot of $P_r(\tau)$ also shows the variation of the radius. The ring widens as it passes over the sphere, and subsequently contracts again. The value of the energy is $E=5.177$ in reduced units.

In Figs. 4 and 5 the ring is initially centered at $Z(0)=-5b$ with momentum $P_r(0)=\rho\Gamma b^2$. The momentum of the sphere is initially $p_0(0)=0$. In Fig. 4 we show the reduced positions $Z(\tau)/b$ and $z_0(\tau)/b$ as functions of $\tau$. In Fig. 5 we show the ring momentum in units $\rho\Gamma b^2$. The ring widens significantly, as it passes over the sphere. The value of the energy is $E=1.150$ in reduced units.

In Figs. 6 and 7 the ring is initially centered at $Z(0)=-5b$ with momentum $P_r(0)=2.5\rho\Gamma b^2$. The momentum of the sphere is initially $p_0(0)=-2.5\rho\Gamma b^2$. In Fig. 6 we show the reduced positions $Z(\tau)/b$ and $z_0(\tau)/b$ as functions of $\tau$. The sphere passes through the ring, and eventually moves to the left. In Fig. 7 we show the momenta in units $\rho\Gamma b^2$. The ring widens as the sphere passes through. The total momentum vanishes at any time. The value of the energy is $E=2.614$ in reduced units.

In Figs. 8 and 9 the ring is initially centered at $Z(0)=5b$ with momentum $P_r(0)=2.5\rho\Gamma b^2$. The momentum of the sphere is initially $p_0(0)=5.4\rho\Gamma b^2$. In Fig. 8 we show the reduced positions $Z(\tau)/b$ and $z_0(\tau)/b$ as functions of $\tau$. The sphere passes through the ring, and eventually moves to the right with a velocity larger than that of the ring. In Fig. 9 we show the momenta in units $\rho\Gamma b^2$. Again the ring widens as the sphere passes through. The value of the energy is $E=4.461$ in reduced units.

In Figs. 10 and 11 the ring is initially centered at $Z(0)=5b$ with momentum $P_r(0)=2.5\rho\Gamma b^2$. The momentum of the sphere is initially $p_0(0)=5\rho\Gamma b^2$. In Fig. 10 we show the reduced positions $Z(\tau)/b$ and $z_0(\tau)/b$ as functions of $\tau$. The sphere catches up with the ring, lingers near the ring, but eventually moves to the right with a velocity smaller than that of the ring. In Fig. 11 we show the momenta in units $\rho\Gamma b^2$. The ring widens as the sphere approaches and keeps the larger width as it moves away. Both the sphere and the ring lose velocity in the encounter. The value of the energy is $E=4.129$ in reduced units.

\section{\label{V}Conclusions}

The coaxial system of vortex ring and sphere has quite complicated dynamics due to the interference of flow patterns. The dynamics is summarized in Hamiltonian equations of motion with Hamiltonian given by Eq. (3.9). The interaction between vortex ring and sphere can be quite intricate, depending on initial conditions and parameters. The dynamics of several coaxial vortex rings and spheres will be even more complicated, but can also be studied on the basis of the corresponding Hamiltonian.

Acoustic waves emitted by a vortex ring interacting with a fixed solid sphere were studied experimentally and theoretically by Minota et al. \cite{14}. The sound radiation is calculated from the rate of change of the scalar potential \cite{15},\cite{16}.

Coaxial interactions of two vortex rings or of a ring with a body were studied numerically by Ye et al. \cite{17}. The interaction of a vortex ring and a fixed rigid sphere was studied in computer simulation by Ryu and Lee \cite{18}. These authors considered also off-axis collisions.

It would be of interest to study the interactions between a vortex ring and a moving sphere in experiment and computer simulation. The present theory provides the tools to predict behavior and allows selection of interesting cases.

The Hamiltonian in Eq. (3.9) has an interest of its own. From a theoretical point of view it may be tempting to study the corresponding quantummmechanical problem.

We have considered the inviscid limit. The effect of viscosity on the behavior of a single vortex ring was discussed in detail by Fukumoto and Moffatt \cite {19}. The effect of viscosity on the interactions between a ring and a rigid sphere, and the absorption of a ring by a sphere are topics demanding investigation.

\newpage

\section*{Figure captions}
\subsection*{Fig. 1}
Sketch of the geometry of colliding vortex ring and sphere. Both are centered on the $z$ axis.

\subsection*{Fig. 2}
Plot of the position of the ring $Z(\tau)/b$ (solid curve) and of the center of the sphere $z_0(\tau)/b$ (dashed curve) as functions of dimensionless time $\tau=\Gamma t/b^2$ for initial conditions $Z(0)=-5b,\;z_0(0)=0,\;P_r(0)=10\rho\Gamma b^2,\;p_0(0)=0$.

\subsection*{Fig. 3}
Plot of the momentum of the ring $P_r(\tau)/(\rho\Gamma b^2)$ as a function of dimensionless time $\tau=\Gamma t/b^2$ for initial conditions as in Fig. 2.

\subsection*{Fig. 4}
Plot of the position of the ring $Z(\tau)/b$ (solid curve) and of the center of the sphere $z_0(\tau)/b$ (dashed curve) as functions of dimensionless time $\tau=\Gamma t/b^2$ for initial conditions $Z(0)=-5b,\;z_0(0)=0,\;P_r(0)=\rho\Gamma b^2,\;p_0(0)=0$.

\subsection*{Fig. 5}
Plot of the momentum of the ring $P_r(\tau)/(\rho\Gamma b^2)$ as a function of dimensionless time $\tau=\Gamma t/b^2$ for initial conditions as in Fig. 4.

\subsection*{Fig. 6}
Plot of the position of the ring $Z(\tau)/b$ (solid curve) and of the center of the sphere $z_0(\tau)/b$ (dashed curve) as functions of dimensionless time $\tau=\Gamma t/b^2$ for initial conditions $Z(0)=-5b,\;z_0(0)=0,\;P_r(0)=2.5\rho\Gamma b^2,\;p_0(0)=-2.5\rho\Gamma b^2$.

\subsection*{Fig. 7}
Plot of the momentum of the ring $P_r(\tau)/(\rho\Gamma b^2)$ (solid curve) and of the sphere momentum $p_0(\tau)/(\rho\Gamma b^2)$ (dashed curve) as functions of dimensionless time $\tau=\Gamma t/b^2$ for initial conditions as in Fig. 6.

\subsection*{Fig. 8}
Plot of the position of the ring $Z(\tau)/b$ (solid curve) and of the center of the sphere $z_0(\tau)/b$ (dashed curve) as functions of dimensionless time $\tau=\Gamma t/b^2$ for initial conditions $Z(0)=5b,\;z_0(0)=0,\;P_r(0)=2.5\rho\Gamma b^2,\;p_0(0)=5.4\rho\Gamma b^2$.

\subsection*{Fig. 9}
Plot of the momentum of the ring $P_r(\tau)/(\rho\Gamma b^2)$ (solid curve) and of the sphere momentum $p_0(\tau)/(\rho\Gamma b^2)$ (dashed curve) as functions of dimensionless time $\tau=\Gamma t/b^2$ for initial conditions as in Fig. 8.

\subsection*{Fig. 10}
Plot of the position of the ring $Z(\tau)/b$ (solid curve) and of the center of the sphere $z_0(\tau)/b$ (dashed curve) as functions of dimensionless time $\tau=\Gamma t/b^2$ for initial conditions $Z(0)=5b,\;z_0(0)=0,\;P_r(0)=2.5\rho\Gamma b^2,\;p_0(0)=5\rho\Gamma b^2$.

\subsection*{Fig. 11}
Plot of the momentum of the ring $P_r(\tau)/(\rho\Gamma b^2)$ (solid curve) and of the momentum of the sphere $p_0(\tau)/(\rho\Gamma b^2)$ (dashed curve) as functions of dimensionless time $\tau=\Gamma t/b^2$ for initial conditions as in Fig. 10.

\newpage
\clearpage
\newpage
\setlength{\unitlength}{1cm}
\begin{figure}
 \includegraphics{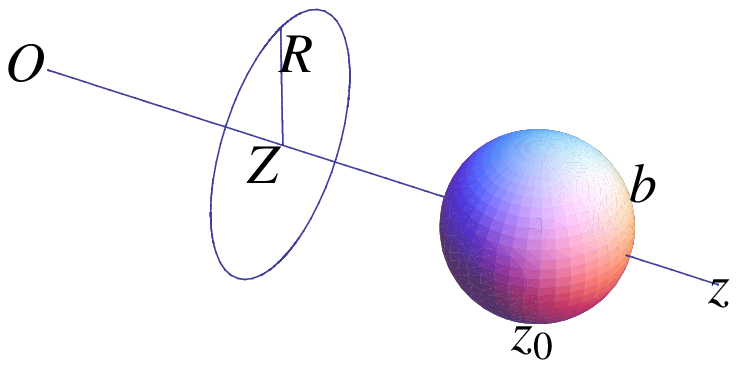}
   \put(-9.1,3.1){}
\put(-1.2,-.2){}
  \caption{}
\end{figure}
\newpage
\clearpage
\newpage
\setlength{\unitlength}{1cm}
\begin{figure}
 \includegraphics{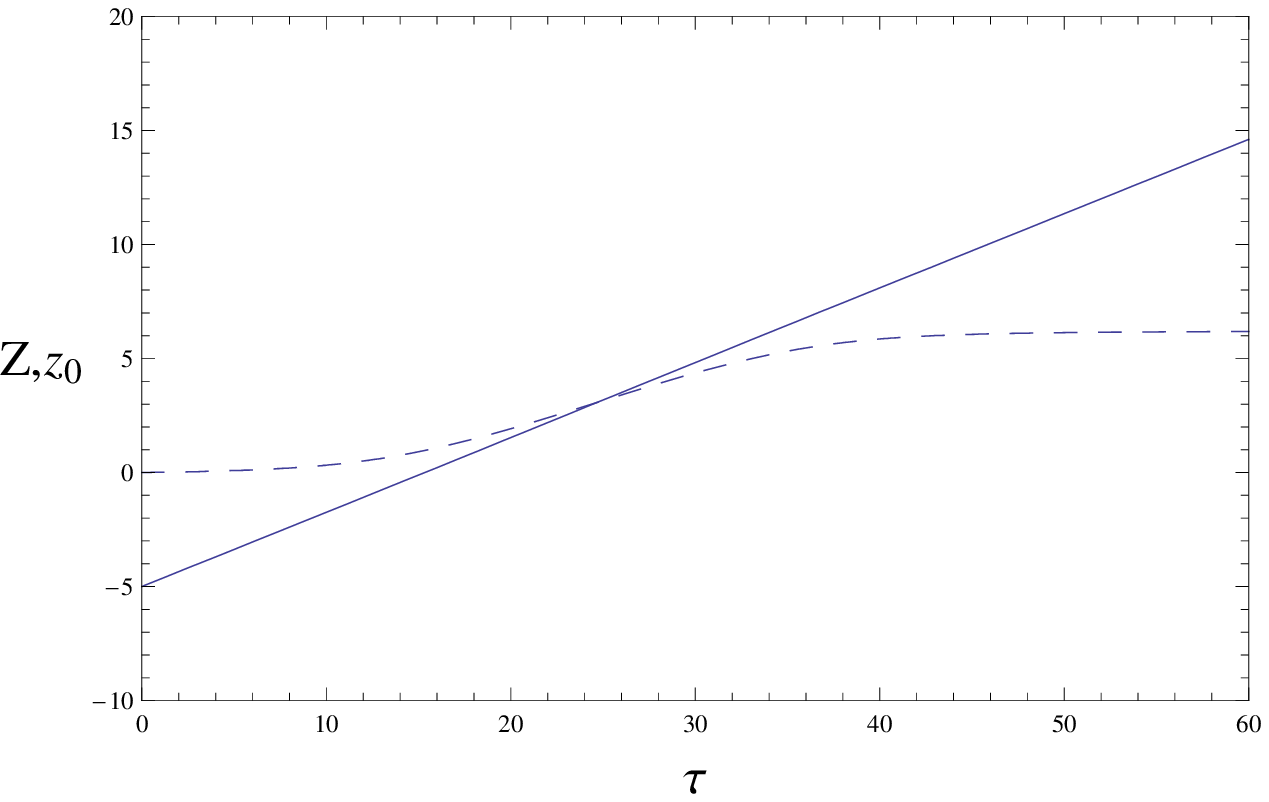}
   \put(-9.1,3.1){}
\put(-1.2,-.2){}
  \caption{}
\end{figure}
\newpage
\clearpage
\newpage
\setlength{\unitlength}{1cm}
\begin{figure}
 \includegraphics{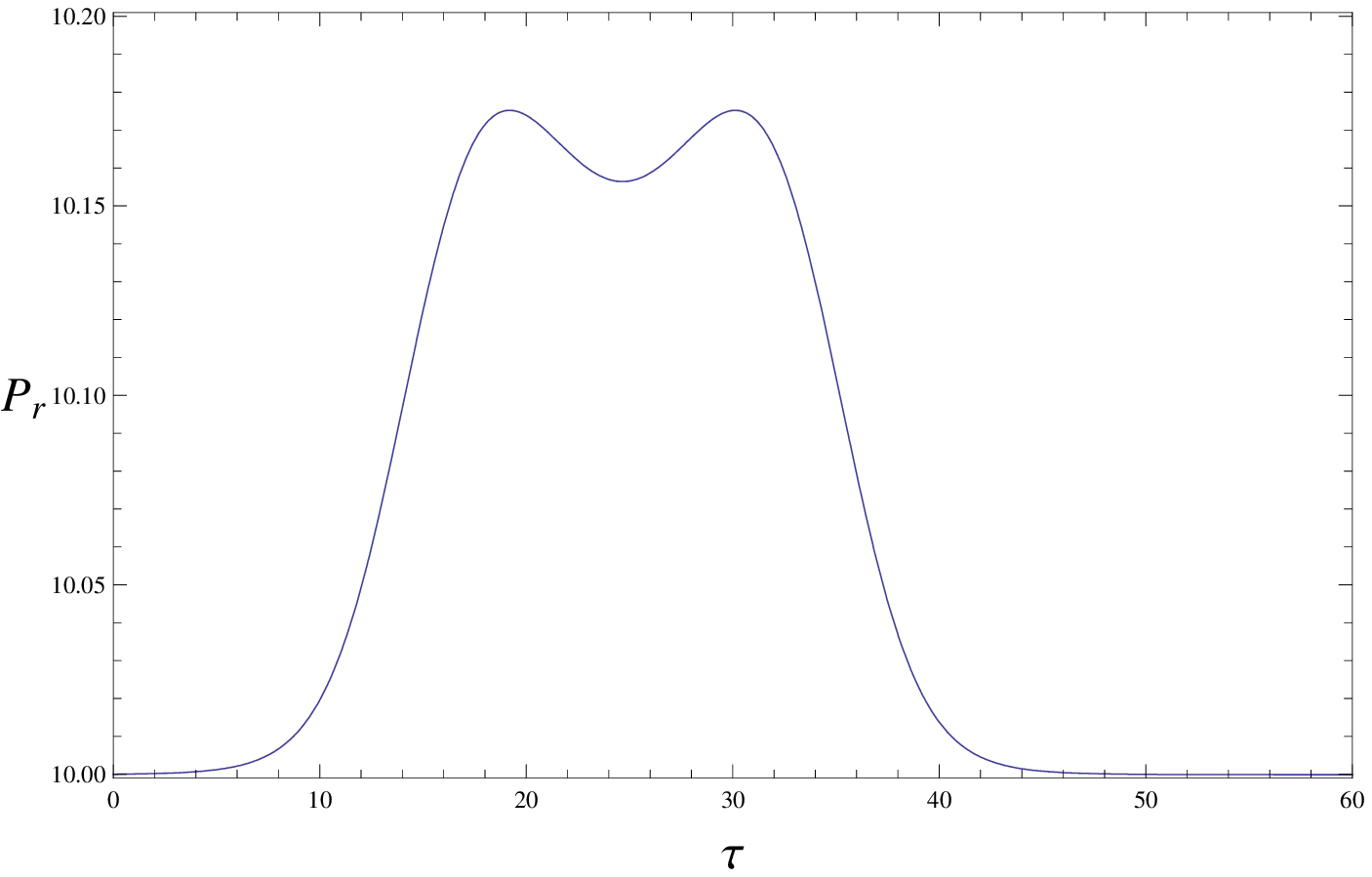}
   \put(-9.1,3.1){}
\put(-1.2,-.2){}
  \caption{}
\end{figure}
\newpage
\clearpage
\newpage
\setlength{\unitlength}{1cm}
\begin{figure}
 \includegraphics{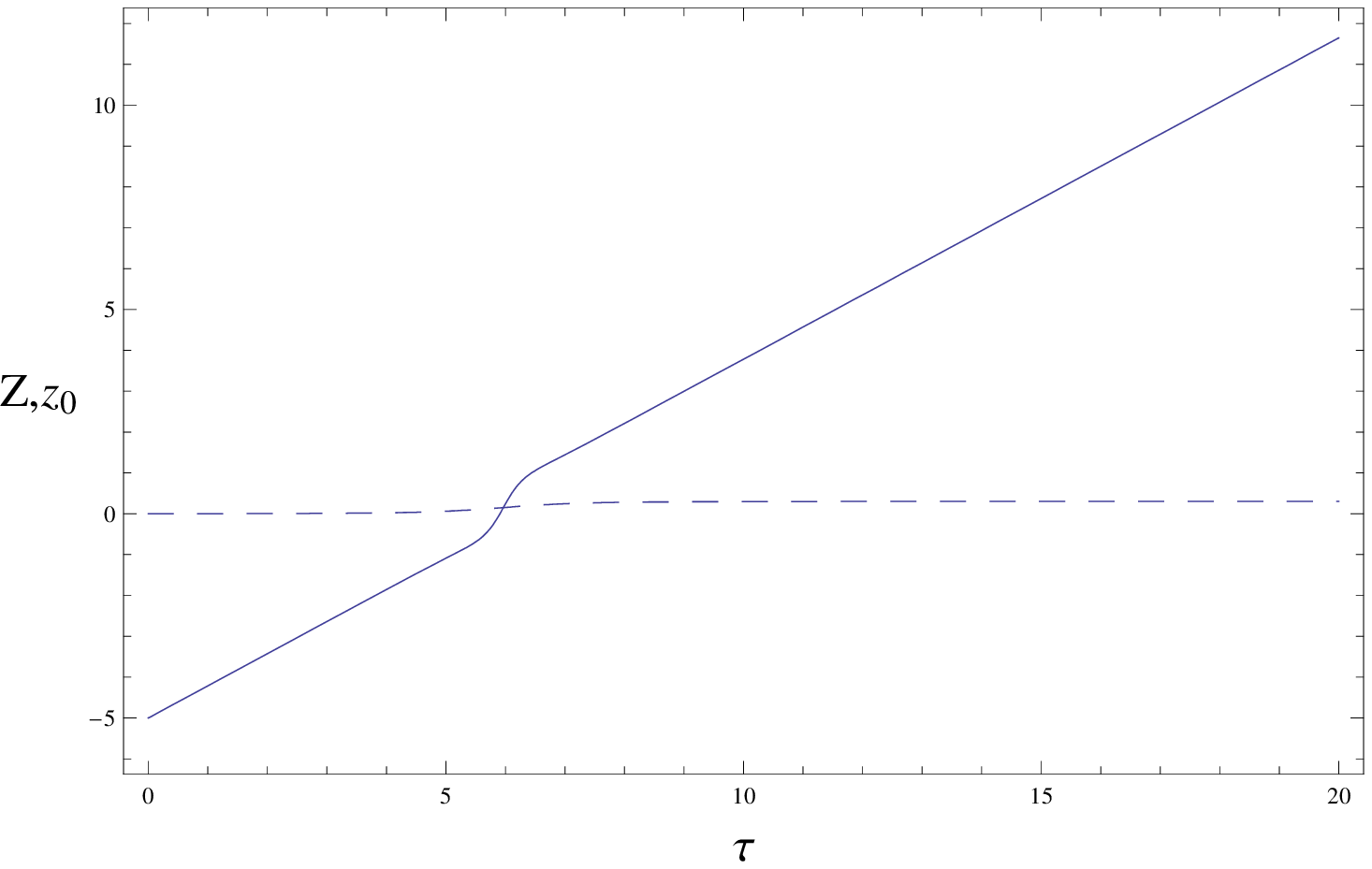}
   \put(-9.1,3.1){}
\put(-1.2,-.2){}
  \caption{}
\end{figure}
\newpage
\clearpage
\newpage
\setlength{\unitlength}{1cm}
\begin{figure}
 \includegraphics{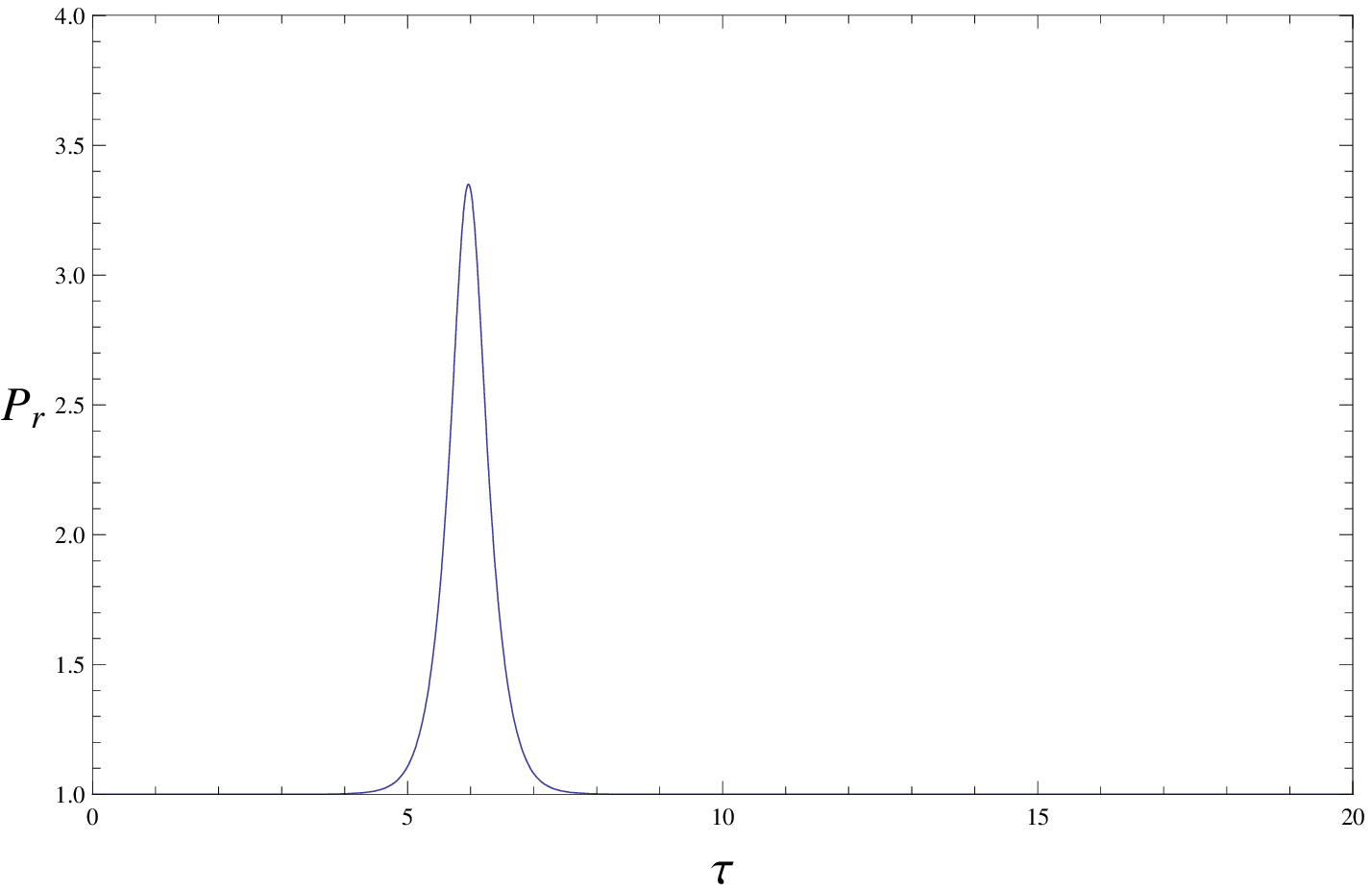}
   \put(-9.1,3.1){}
\put(-1.2,-.2){}
  \caption{}
\end{figure}
\newpage
\clearpage
\newpage
\setlength{\unitlength}{1cm}
\begin{figure}
 \includegraphics{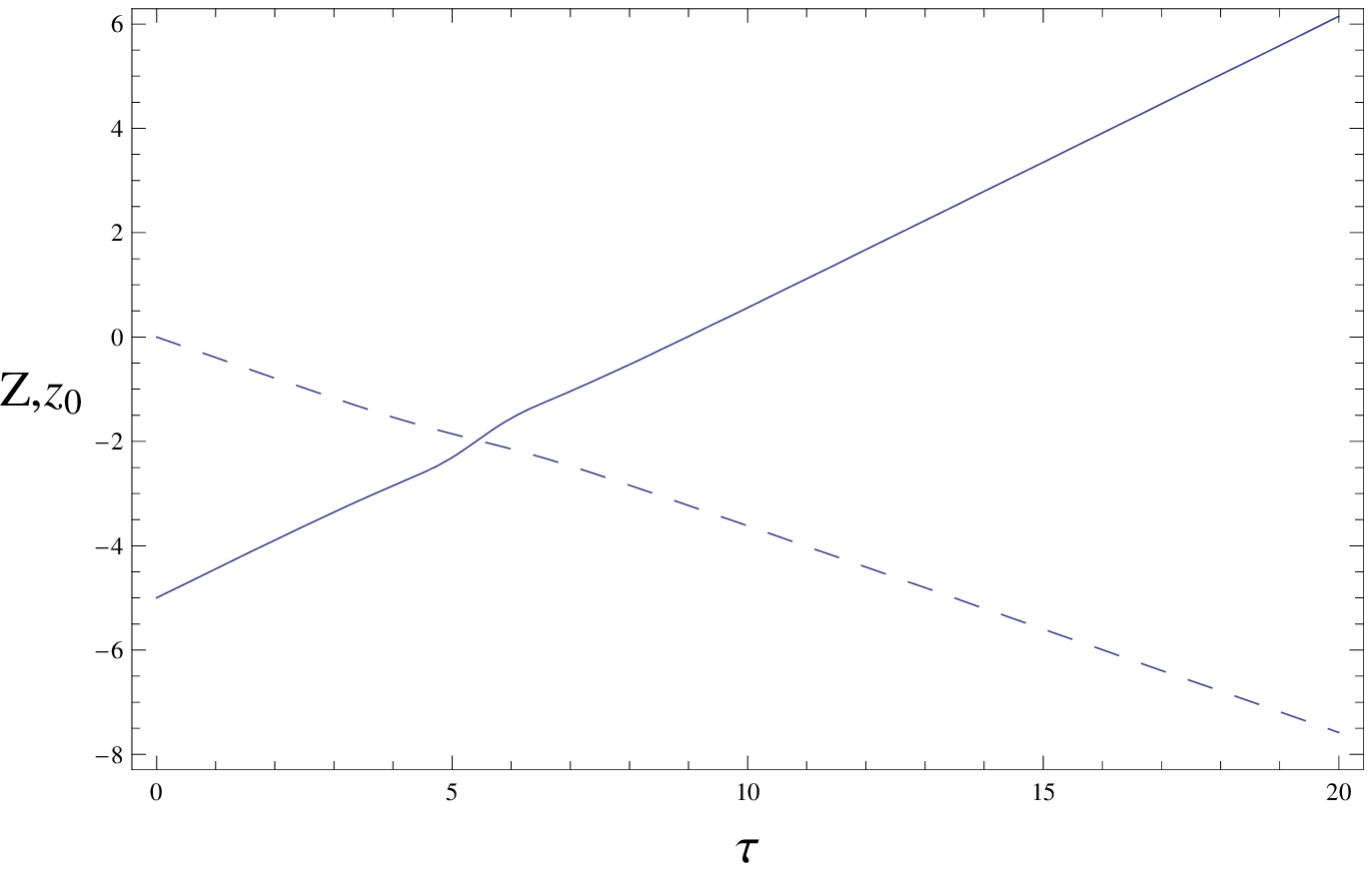}
   \put(-9.1,3.1){}
\put(-1.2,-.2){}
  \caption{}
\end{figure}
\newpage
\clearpage
\newpage
\setlength{\unitlength}{1cm}
\begin{figure}
 \includegraphics{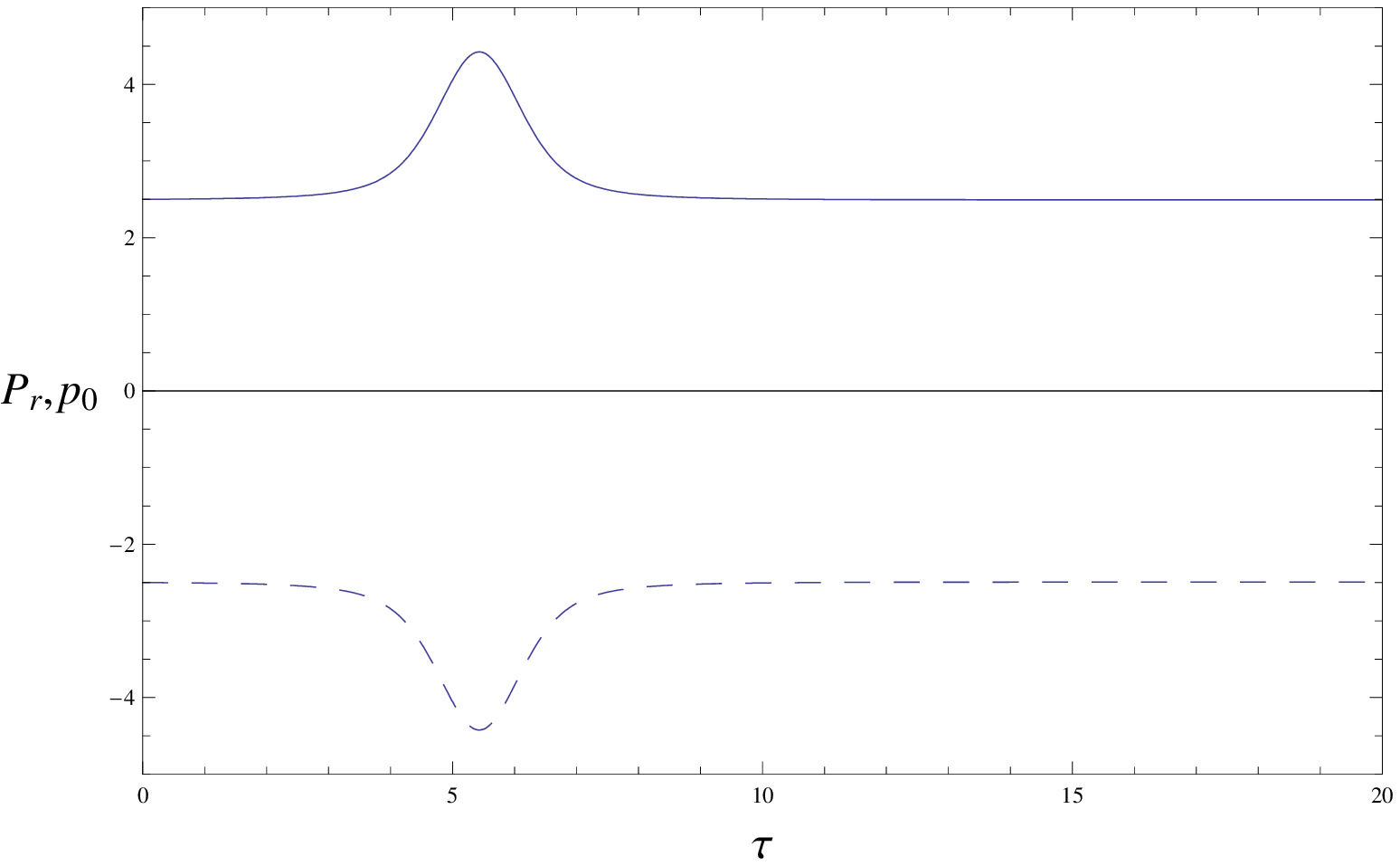}
   \put(-9.1,3.1){}
\put(-1.2,-.2){}
  \caption{}
\end{figure}
\newpage
\clearpage
\newpage
\setlength{\unitlength}{1cm}
\begin{figure}
 \includegraphics{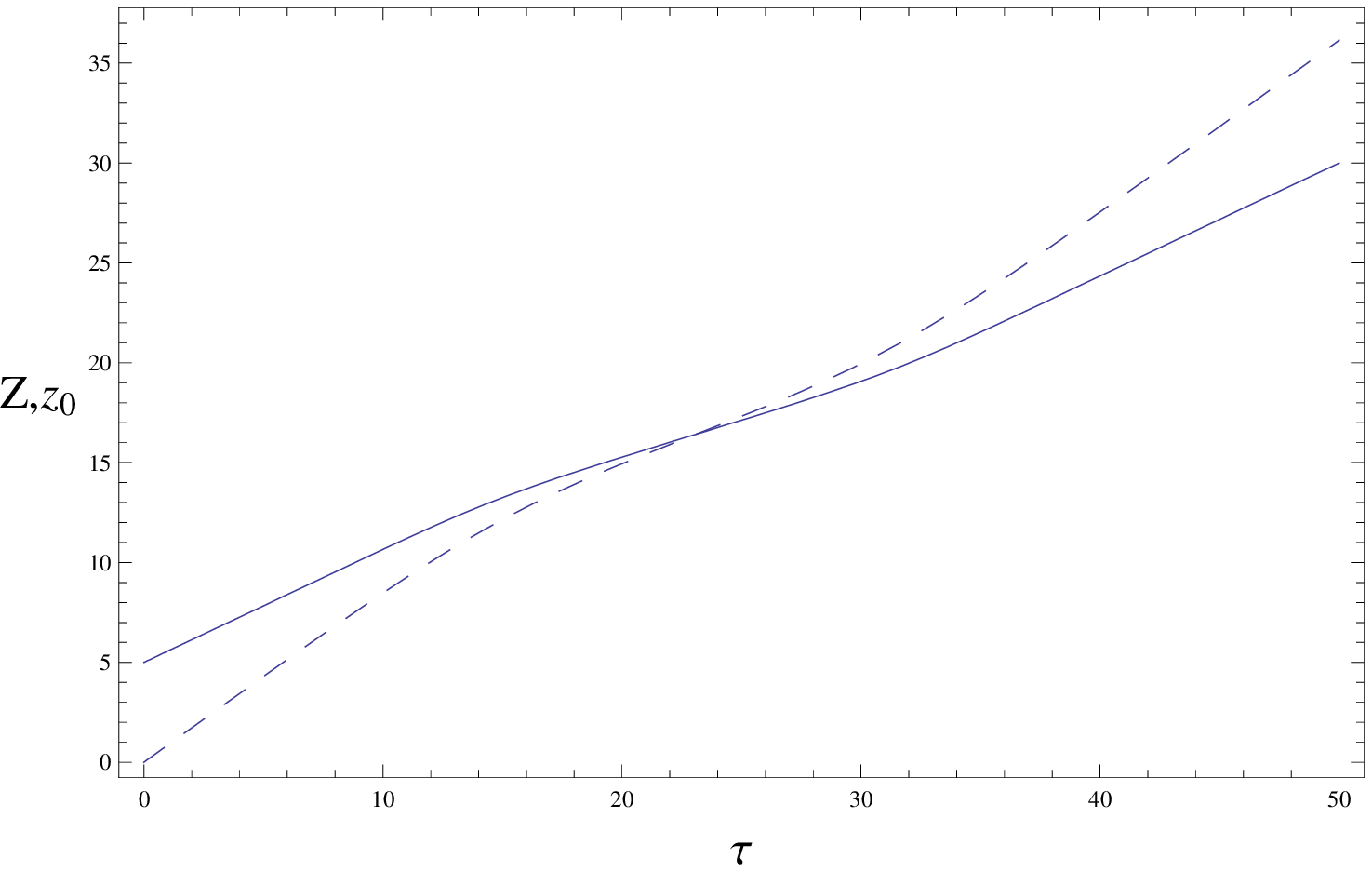}
   \put(-9.1,3.1){}
\put(-1.2,-.2){}
  \caption{}
\end{figure}
\newpage
\clearpage
\newpage
\setlength{\unitlength}{1cm}
\begin{figure}
 \includegraphics{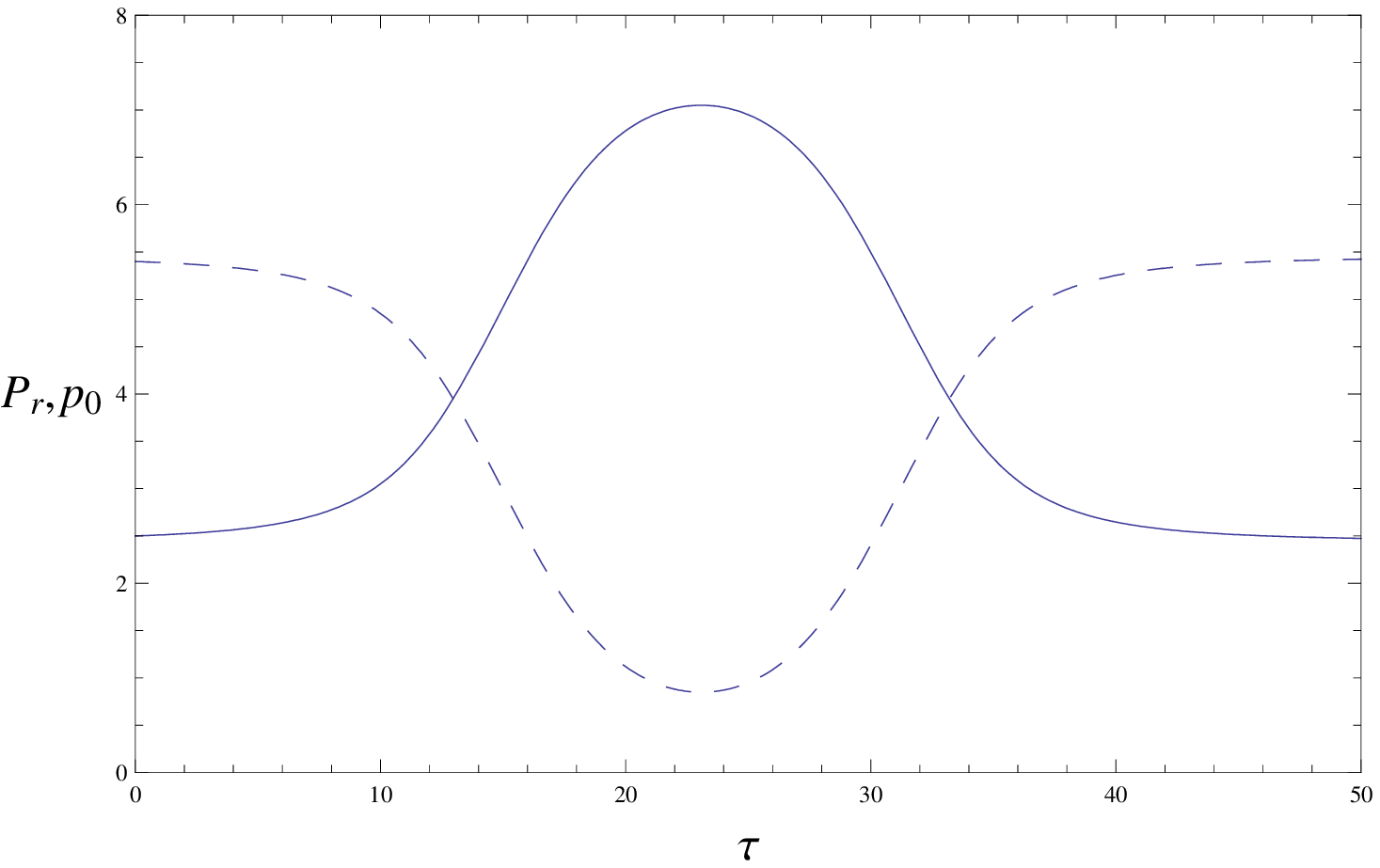}
   \put(-9.1,3.1){}
\put(-1.2,-.2){}
  \caption{}
\end{figure}
\newpage
\clearpage
\newpage
\setlength{\unitlength}{1cm}
\begin{figure}
 \includegraphics{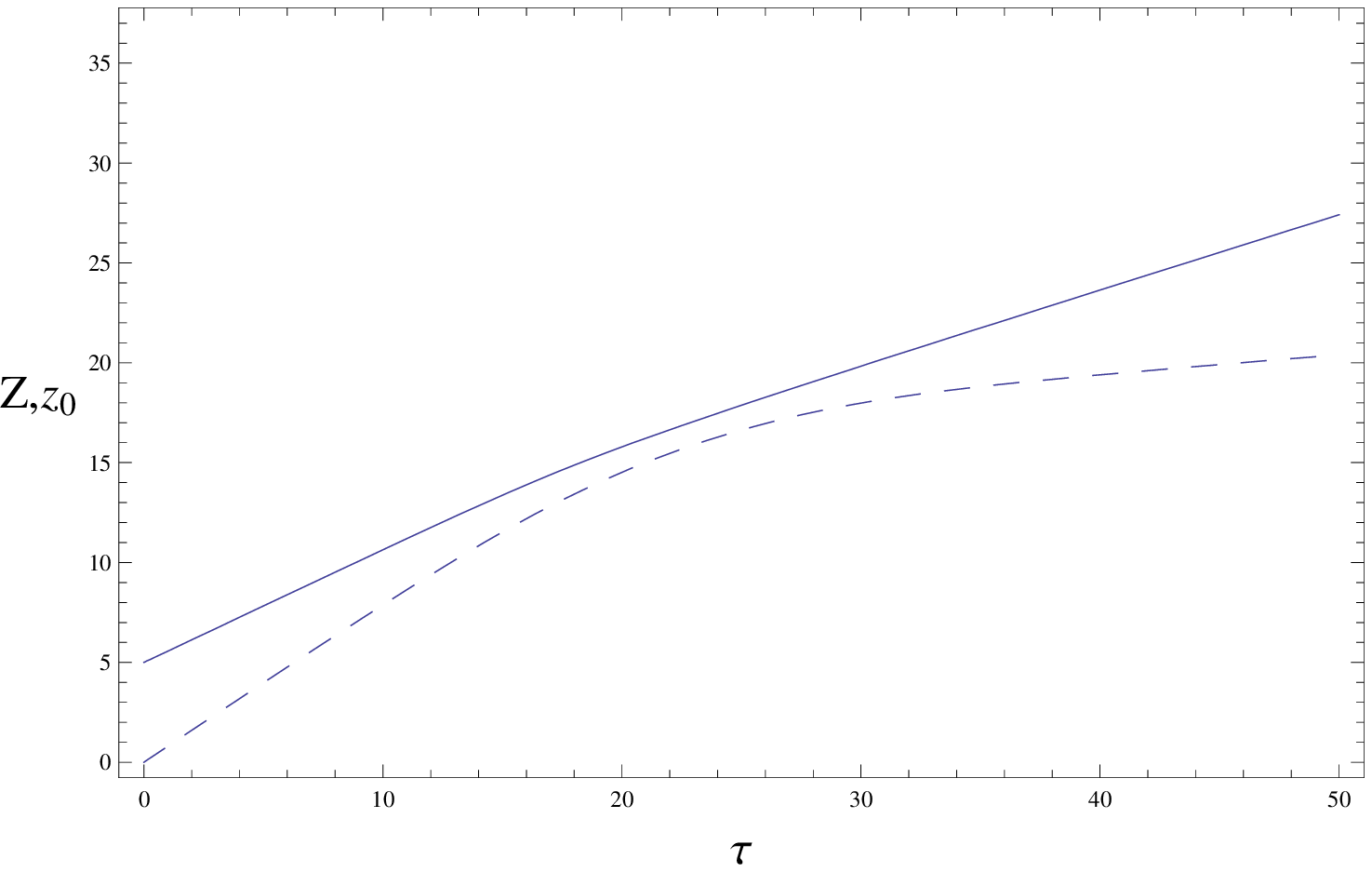}
   \put(-9.1,3.1){}
\put(-1.2,-.2){}
  \caption{}
\end{figure}
\newpage
\clearpage
\newpage
\setlength{\unitlength}{1cm}
\begin{figure}
 \includegraphics{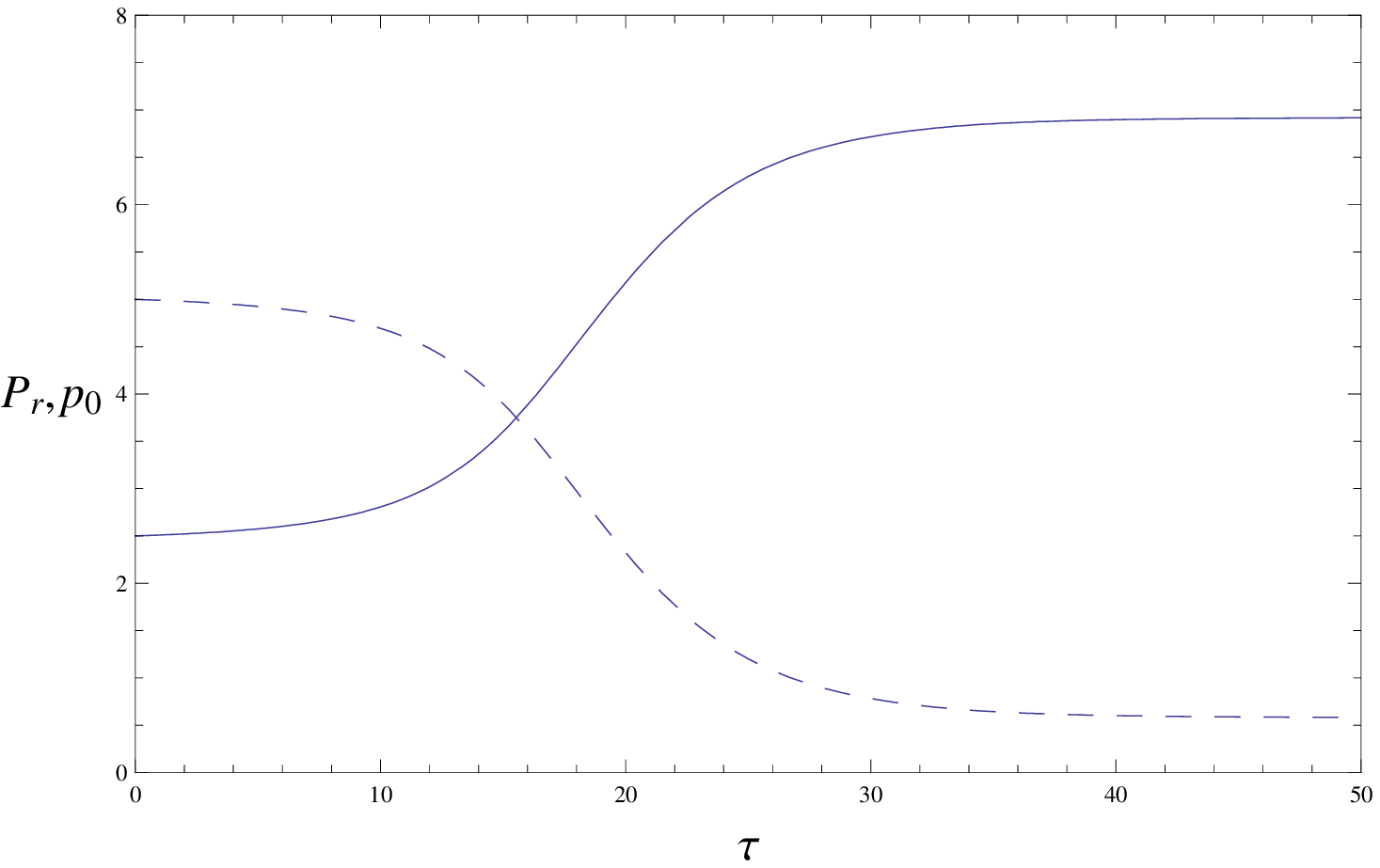}
   \put(-9.1,3.1){}
\put(-1.2,-.2){}
  \caption{}
\end{figure}
\newpage

\end{document}